\documentclass[12pt]{article}
\usepackage{amsmath,amssymb}
\usepackage{bm}
\usepackage{graphicx}
\usepackage{color}
\usepackage{wrapfig}
\begin{document}
	\title{
		\begin{flushright}
			\ \\*[-80pt]
			\begin{minipage}{0.17\linewidth}
				\normalsize
				HUPD1712 \\ \\ \\
			\end{minipage}
		\end{flushright}
		{\Large \bf   Neutrino CP violation and sign of baryon asymmetry in the minimal seesaw model
			\\*[20pt]}}

\author{
\centerline{Yusuke~Shimizu$^{1,}$\footnote{E-mail address: yu-shimizu@hiroshima-u.ac.jp} \ ,
 Kenta~Takagi$^{1,}$\footnote{E-mail address: takagi-kenta@hiroshima-u.ac.jp}  \ ,  and
Morimitsu~Tanimoto$^{2,}$\footnote{E-mail address: tanimoto@muse.sc.niigata-u.ac.jp} }
\\*[20pt]
\centerline{
\begin{minipage}{\linewidth}
\begin{center}
$^1${\it \normalsize
\it \normalsize \it \normalsize Graduate School of Science, Hiroshima University, \\ Higashi -Hiroshima 739-8526, Japan}
\\*[4pt]
$^2${\it \normalsize
Department of Physics, Niigata University,~Niigata 950-2181, Japan }
\end{center}
\end{minipage}}
\\*[70pt]}

\date{
\centerline{\small \bf Abstract}
\begin{minipage}{0.9\linewidth}
\vskip  1 cm
\small
We discuss the correlation between the CP violating Dirac phase of the lepton mixing matrix and the cosmological baryon asymmetry based on the leptogenesis in the minimal seesaw model
with two right-handed Majorana neutrinos and the trimaximal mixing for neutrino flavors.
The sign of the CP violating Dirac phase at low energy is fixed by the observed cosmological baryon asymmetry
since there is only one phase parameter in the model.
According to the recent T2K and NO$\nu$A data of the CP violation,
the Dirac neutrino mass matrix of our model is fixed only for 
the normal hierarchy of neutrino masses.
\end{minipage}
}

\begin{titlepage}
\maketitle
\thispagestyle{empty}
\end{titlepage}

\section{Introduction}

Recent T2K data strongly indicate the CP violation in the neutrino oscillation \cite{Abe:2017vif,T2K}.
The NO$\nu$A data also suggest the CP violation \cite{Adamson:2017gxd}, which is consistent with the T2K result.
We are in the era to develop the flavor structure of Yukawa couplings by focusing on the leptonic CP violation  since the CP violating phase of neutrinos will be observed in the near future.

We study the flavor structure in the seesaw model \cite{Minkowski}-\cite{MSSV} to reveal the underlying physics of the lepton flavors.
For this purpose, it is advantageous to consider the minimum number of parameters in the neutrino mass matrices needed for reproducing the neutrino mixing angles and CP violating phases completely \cite{Shimizu:2012ry}.
There are many attempts toward the minimal seesaw model \cite{King:1998jw}-\cite{Rink:2016knw}.

In our previous work, we studied the minimal seesaw model taking account of the CP violation of neutrinos \cite{Shimizu:2017fgu},
where we assumed two right-handed Majorana neutrinos and took the trimaximal mixing pattern of neutrino flavors ~\cite{Grimus:2008tt,Albright:2008rp}. 
The trimaximal mixing is realized by the non-Abelian discrete flavor symmetry \cite{Altarelli:2010gt}-\cite{King:2014nza}.
Since this mixing pattern is the simple framework obtained by the additional rotation of $2-3$ ($\rm TM_1$) or 
$1-3$ ($\rm TM_2$) families of neutrinos \cite{Albright:2010ap,Rodejohann:2012cf} to the tribimaximal mixing (TBM) basis~\cite{Harrison:2002er,Harrison:2002kp}, 
we could find the relations among mixing angles and CP violating phase, so called mixing sum rules \cite{Antusch:2011ic}-\cite{Petcov:2014laa}.
The relations among  mixing angles and CP violating phase have been discussed intensively \cite{Marzocca:2013cr}-\cite{Morozumi:2017rrg}.
We investigated the flavor structure of the Dirac neutrino mass matrix  in our framework of the minimal seesaw model.
The desirable candidates of Dirac neutrino mass matrices were obtained in the diagonal basis of the charged lepton mass matrix and the $2\times 2$ right-handed Majorana neutrino mass matrix for both cases of the normal hierarchy (NH) and the inverted hierarchy (IH) of neutrino masses, respectively.
However, we could not determine the sign of the CP violating phase $\delta_{CP}$ \cite{Shimizu:2017fgu}.

As well known, it is possible to correlate the CP violation at the low energy with the CP violation at the high energy \cite{Endoh:2002wm,Pascoli:2006ci} through the leptogenesis \cite{Fukugita:1986hr}.
In this work, it is found that
the CP violating phase in the minimal seesaw model is directly related to the baryon asymmetry of the universe (BAU).
We discuss the correlation between the predicted CP violating phase $\delta_{CP}$ and  BAU through the leptogenesis in our minimal seesaw model.
Finally, we determine the flavor structure of the Dirac neutrino mass matrix so that it  explains both the low energy neutrino experimental data and BAU.

The paper is organized as follows.
We summarize our minimal seesaw model \cite{Shimizu:2017fgu} in section 2,
where the structure of the Dirac neutrino mass matrix is discussed to reproduce $\rm TM_1$ or $\rm TM_2$ in NH or IH.
In section 3, we present the framework of the leptogenesis in our minimal seesaw model.
In section 4, the correlation between the low energy CP violation and BAU is discussed.
The section 5 is devoted to the summary and discussions.
Appendix A gives the detailed studies of the Dirac neutrino mass matrices and
Appendix B presents the explicit form of the relevant mixing matrix elements in our model.

\section{Our minimal  seesaw model}

Our minimal seesaw model consists of two right-handed Majorana neutrinos $N_1$ and $N_2$, 
and three left-handed neutrinos in Type I seesaw \cite{Shimizu:2012ry}.
The $3\times 2$ Dirac neutrino mass matrix is constrained by a certain principle
in the diagonal basis of both the charged lepton and right-handed Majorana neutrino mass matrices.
We have already proposed simple Dirac neutrino mass matrices based on the symmetry of the neutrino mixing matrix, $\rm TM_1$ and $\rm TM_2$.
We summarize them in the following subsections.
\subsection{$\rm TM_1$: 2-3 family mixing in NH}
For the case of $\rm TM_1$ in the NH neutrino masses,
the Dirac neutrino mass matrix $M_D$ is given in terms of the Yukawa matrix $Y_\nu$ in the basis of the diagonal right-handed Majorana neutrino mass matrix $M_R$ as follows~\cite{Shimizu:2017fgu}:
 \begin{equation}
 M_D=v Y_\nu=v
 \begin{pmatrix}
 \frac{b+c}{2} & \frac{e+f}{2} \\
 b & e \\
 c & f
 \end{pmatrix}, \qquad\qquad
 M_R=
 \begin{pmatrix}
 M_1
 & 0 \\
 0 & M_2
 \end{pmatrix}
 =M_0
 \begin{pmatrix}
 p^{-1}
 & 0 \\
 0 & 1
 \end{pmatrix},
 \label{MDgeneral}
 \end{equation}
where $v=174.1$GeV, $M_0$ is the mass scale of the right-handed Majorana neutrino and $p$ is the ratio $M_2/M_1$.
We take $e$ and $f$ to be real and $b$ and $c$ to be complex in general by using the freedom of redefinitions of phases in the left-handed lepton fields.
By using the type I seesaw, we obtain the left-handed neutrino mass matrix $M_\nu$:
\begin{equation}
M_{\nu }=-M_D M_R^{-1}M_D^T \ ,
\end{equation}
which turns  to
\begin{equation}
\hat M_\nu= V_{\text{TBM}}^T M_\nu V_{\text{TBM}} = -\frac{v^2}{M_0}
\begin{pmatrix}
0 & 0 & 0 \\
0 & \frac{3}{4}\left ((b+c)^2p+(e+f)^2\right ) &
\frac{1}{2}\sqrt{\frac{3}{2}}\left ((c^2-b^2)p-e^2+f^2\right ) \\
0 & \frac{1}{2}\sqrt{\frac{3}{2}}\left ((c^2-b^2)p-e^2+f^2\right ) &
\frac{1}{2}\left ((b-c)^2p+(e-f)^2\right )
\end{pmatrix},
\label{TM1NH}
\end{equation}
where
\begin{equation}
V_{\text{TBM}}=
\begin{pmatrix}
\frac{2}{\sqrt{6}} & \frac{1}{\sqrt{3}} & 0 \\
-\frac{1}{\sqrt{6}} & \frac{1}{\sqrt{3}} & -\frac{1}{\sqrt{2}} \\
-\frac{1}{\sqrt{6}} & \frac{1}{\sqrt{3}} & \frac{1}{\sqrt{2}}
\end{pmatrix}.
\label{UTBM}
\end{equation}
This mass matrix $ \hat M_\nu $ is given in the TBM mixing basis, which is derived  in Appendix A.
The neutrino mass matrix in Eq.~(\ref{TM1NH}) is diagonalized by the rotation of $2-3$ families as
\begin{equation}
V_{23}=\frac{1}{{\cal A}}
\begin{pmatrix}
{\cal A} & 0 & 0 \\
0 & 1& {\cal V}\\
0 &  -{\cal V}^* & 1
\end{pmatrix}, \qquad\qquad {\cal A}=\sqrt{1+|{\cal V}|^2} \ ,
\label{rotaion23}
\end{equation}
where ${\cal V}$ is given in terms of $b$, $c$, $e$, $f$ and $p$.
The PMNS matrix \cite{Maki:1962mu,Pontecorvo:1967fh}
 is expressed as
\begin{align}
U_\text{PMNS}= V_{\text{TBM}} V_{23} \ ,
\label{MNS}
\end{align}
which gives three mixing angles, one Dirac phase, and one Majorana phase.


In order to remove one complex parameter from the Dirac neutrino mass matrix,
we put one zero in the first column of  the Dirac neutrino mass matrix
 (see Appendix A.1).
There are three cases with one zero.
For case I $(b+c=0)$, the Dirac neutrino mass matrix $M_D$ and 
the left-handed neutrino mass matrix  $\hat M_{\nu }$ are given as
\begin{equation}
{\rm Case\ I}\qquad  b+c=0 \ : \qquad
M_D=v Y_\nu=v
\begin{pmatrix}
0 & \frac{e+f}{2} \\
b & e \\
-b & f
\end{pmatrix},
\label{MD1}
\end{equation}
\begin{equation}
\hat M_{\nu }= -\frac{f^2 v^2}{M_0}
\begin{pmatrix}
0 & 0 & 0 \\
0 & \frac{3}{4}(k+1)^2 & -\frac{1}{2}\sqrt{\frac{3}{2}}(k^2-1) \\
0 & -\frac{1}{2}\sqrt{\frac{3}{2}}(k^2-1) & 2B^2 p e^{2i\phi_B}+\frac{1}{2}(k-1)^2
\end{pmatrix},
\label{case1}
\end{equation}
respectively, where the CP violating phase $\phi_B$ appears in the parameter $b$ such as
\begin{equation}
\frac{e}{f}=k \ , \qquad
  \arg[b]=\phi_B\ , \qquad \frac{b}{f}=B e^{i\phi_B}\ ,
\label{parameters}
\end{equation}
with  $k$ and  $B$ being real.
For case II $(c=0)$, they are
\begin{equation}
{\rm Case\ II}\qquad c=0: \  \qquad M_D=v Y_\nu=v
\begin{pmatrix}
\frac{b}{2} & \frac{e+f}{2} \\
b & e \\
0 & f
\end{pmatrix},
\label{MD2}
\end{equation}
\begin{equation}
\hat M_{\nu }= -\frac{f^2 v^2}{M_0}
\begin{pmatrix}
0 & 0 & 0 \\
0 & \frac{3}{4}[\hat B^2 p e^{2i\phi_B}+(k+1)^2] &
 -\frac{1}{2}\sqrt{\frac{3}{2}}[\hat B^2 p e^{2i\phi_B}+k^2-1] \\
0 &  -\frac{1}{2}\sqrt{\frac{3}{2}}[\hat B^2 p e^{2i\phi_B}+k^2-1] &
\frac{1}{2}[\hat B^2 p e^{2i\phi_B}+(k-1)^2]
\end{pmatrix},
\label{case2}
\end{equation}
where
\begin{equation}
\frac{e}{f}=k \ , \qquad \arg[b]=\phi_B\ , \qquad
\frac{b}{f}=\hat B e^{i\phi_B}\ ,
\label{parameters2}
\end{equation}
with  $k$ and  $\hat B$ being real.
For case III $(b=0)$, they are
\begin{equation}
{\rm Case\ III}\qquad b=0 : \qquad M_D=v Y_\nu=v
\begin{pmatrix}
\frac{c}{2} & \frac{e+f}{2} \\
0 & e \\
c & f
\end{pmatrix},
\label{MD3}
\end{equation}
\begin{equation}
\hat M_{\nu }= - \frac{f^2 v^2}{M_0}
\begin{pmatrix}
0 & 0 & 0 \\
0 & \frac{3}{4}[B^2 p e^{2i\phi_B}+(k+1)^2] &
-\frac{1}{2}\sqrt{\frac{3}{2}}[-B^2 p e^{2i\phi_B}+k^2-1] \\
0 &  -\frac{1}{2}\sqrt{\frac{3}{2}}[-B^2 p e^{2i\phi_B}+k^2-1] &
 \frac{1}{2}[B^2 p e^{2i\phi_B}+(k-1)^2]
\end{pmatrix},
\label{case3}
\end{equation}
where
\begin{equation}
\frac{e}{f}=k \ , \qquad
\arg[c]=\phi_B\ , \qquad \frac{c}{f}=B e^{i\phi_B}\ .
\end{equation}
For three cases,
neutrino masses are given  follows:
\begin{align}
&\phantom{========}m_1=0, \qquad \quad m_2^2 m_3^2=\frac{9 v^8}{4M_0^4}(j-k)^4 f^8 B^4p^2  ,\nonumber \\
m_2^2+m_3^2&=\frac{v^4 f^4}{16M_0^2}\left [B^4p^2(5j^2+2j+5)^2+2B^2p(5jk+j+k+5)^2\cos 2\phi_B+(5k^2+2k+5)^2\right ]
\label{masses}
\end{align}
where $j\equiv b/c=-1$ and $j=0$ for cases I and III, respectively.
For case II, $Bj\equiv\hat B$ with $j\rightarrow -\infty$ and $B\rightarrow 0$.
It is noticed that  the PMNS matrix elements are correlated with neutrino masses.

The explicit form of the $2-3$ flavor mixing ${\cal V}$ in Eq.~(\ref{rotaion23})  is given in Appendix B for the three cases.
We can predict the CP violating measure, the Jarlskog invariant $J_{CP}$ ~\cite{Jarlskog:1985ht}, 
with  $\cal V$'s as follows:
\begin{eqnarray} 
J_{CP}\equiv \text{Im}\left [U_{e1}U_{\mu 2}U_{e2}^\ast U_{\mu 1}^\ast \right ]=-\frac{1}{3\sqrt{6} {\cal A}^2} {\rm Im[{\cal V^*}]} \ ,
\label{J}
\end{eqnarray}
where $U_{\alpha i}$ denotes the PMNS matrix elements of Eq.~(\ref{MNS}).


It is noted that  the Littlest seesaw model
by King {\it et al.}~\cite{King:2015dvf}-\cite{King:2016yvg}
 corresponds to $k=-3$ in case I, that is,
\begin{equation}
M_D=v Y_\nu=v
\begin{pmatrix}
0 & f \\
b & 3f \\
-b & -f
\end{pmatrix}.
\label{King}
\end{equation}

\subsection{$\rm TM_1$: $2-3$ family mixing in IH}

The Dirac neutrino mass matrix $M_D$ is presented in terms of the Yukawa matrix $Y_\nu$ for the case of $\rm TM_1$ in the IH neutrino masses with $m_3=0$.
As shown in Appendix A.2, the Dirac neutrino mass matrix  and the  left-handed Majorana neutrino mass matrices are given as:
\begin{equation}
 M_D=v Y_\nu=v
\begin{pmatrix}
-2b & \frac{e+f}{2} \\
b & e \\
b & f
\end{pmatrix},
\label{MD23-IH}
\end{equation}
and
\begin{equation}
\hat M_{\nu }= -\frac{v^2}{M_0}
\begin{pmatrix}
6b^2 p & 0 & 0 \\
0 & \frac{3}{4}(e+f)^2 & -\frac{1}{2}\sqrt{\frac{3}{2}}(e-f)(e+f) \\
0 & -\frac{1}{2}\sqrt{\frac{3}{2}}(e-f)(e+f) & \frac{1}{2}(e-f)^2
\end{pmatrix}
,
\end{equation}
respectively, where $m_1=6b^2p v^2/M_0$ and $m_3=0$.
One can take the elements in the first column of the Dirac neutrino mass matrix real with the freedom of redefinitions of phases, and 
 $e$ and $f$ are complex.
The relative phase between $e$ and $f$ leads to the CP violation in contrast to the case of subsection 2.1.

\subsection{$\rm TM_2$: $1-3$ family mixing in NH or IH}

Finally in section 2, we present the Dirac neutrino mass matrix $M_D$ in terms of the Yukawa matrix $Y_\nu$ in the case of $\rm TM_2$ for both NH and  IH neutrino masses.
As seen in Appendix A.3, the Dirac neutrino mass matrix and the left-handed Majorana neutrino mass matrix are  given as follows:
\begin{equation}
 M_D=v Y_\nu=v
\begin{pmatrix}
b & -e-f \\
b & e \\
b& f
\end{pmatrix},
\label{MD13}
\end{equation}
\begin{equation}
\hat M_\nu = -\frac{v^2}{M_0}
\begin{pmatrix}
\frac{3}{2}(e+f)^2 & 0 &\frac{\sqrt{3}}{2}(e^2-f^2)\\
0 & 3b^2 p & 0 \\
\frac{\sqrt{3}}{2}(e^2-f^2) &0 & \frac{1}{2}(e-f)^2
\end{pmatrix},
\label{Mnu13}
\end{equation}
respectively, where $m_2=3b^2pv^2/M_0$ with $m_1=0$ or $m_3=0$.


\section{Implications on Leptogenesis}

Our minimal seesaw model predicts the magnitude of the CP violating phase  $\delta_{CP}$.
This result encourages us to discuss a cosmological consequence of the type-I seesaw mechanism.
The CP violating decays of the heavy Majorana neutrinos can explain the observed BAU by the mechanism of the leptogenesis \cite{Fukugita:1986hr}.
Since our seesaw model has only one phase parameter $\phi_B$,
we expect that the sign of the lepton asymmetry of the  leptogenesis is related with the sign of $\delta_{CP}$ \cite{Frampton:2002qc,Bhattacharya:2006aw,Harigaya:2012bw,
 Bambhaniya:2016rbb,Rink:2016knw,Endoh:2002wm}.

Let us discuss the numerical implications of our models  on  the leptogenesis.
We assume the lightest right-handed neutrino $N_1$ is much lighter than $N_2$ for simplicity.
This condition suffices for the flavor independent analysis.
The flavor summed CP asymmetry at the decay of the lighter right-handed neutrino $N_1$ is given as \cite{Covi:1996wh,Buchmuller:2000as,Giudice:2003jh}
\begin{equation}
\epsilon_{N_1}=-\frac{1}{8\pi} \sum_j
\frac{{\rm Im}[\{(Y_\nu^\dagger Y_\nu)_{j1}\}^2]}{(Y_\nu^\dagger Y_\nu)_{11}}
\left [f^V \left (\frac{M_j^2}{M_1^2}\right )+
f^S \left (\frac{M_j^2}{M_1^2}\right )\right ] \ ,
\label{asym}
\end{equation}
where $f^V(x)$ and  $f^S(x)$  are the contributions from vertex and self-energy corrections, respectively.
In the case of the standard model (SM) with right-handed neutrinos,
they are given as
\begin{equation}
f^V(x)=\sqrt{x}\left [ (x+1)\ln\left ( 1+\frac{1}{x}\right )-1\right ], \quad\quad
f^S(x)=\frac{\sqrt{x}}{x-1} .
\end{equation}
For the case of $p=M_2/M_1\gg 1$,  the CP asymmetry is approximately expressed as
\begin{equation}
\epsilon_{N_1}\simeq -\frac{3}{16\pi} \sum_j
\frac{{\rm Im}[\{(Y_\nu^\dagger Y_\nu)_{j1}\}^2]}{(Y_\nu^\dagger Y_\nu)_{11}}
\frac{1}{p}\ .
\label{approasym}
\end{equation}
The Yukawa matrix $Y_\nu$ is given as the $3\times 2$ matrix as discussed in the section 2.

By  assuming no pre-existing asymmetry of $N_1$,
the final amount of $B-L$ asymmetry $Y_{B-L}$ can be conveniently written as
\begin{equation}
Y_{B-L}=-\epsilon_{N_1} \kappa Y_{N_1}^{eq}(T\gg M_1) \ ,
\label{B-L}
\end{equation}
where
\begin{equation}
Y_{N_1}^{eq}(T\gg M_1)=\frac{135 \ \zeta(3)}{4\pi g^*} \ ,
\end{equation}
with $g*=106.75$ for SM.
The parameter $\kappa$ is a suppression factor which accounts for
the number density of $N_1$ with respect to the equilibrium value, the out-of-equilibrium condition at the decay, and the thermal corrections.
It is approximately given as \cite{Giudice:2003jh}
\begin{equation}
\frac{1}{\kappa}\simeq \frac{3.3\times 10^{-3}}{\tilde m_1}
+\left (\frac{\tilde m_1}{5.5\times 10^{-4} {\rm eV}} \right )^{1.16} \ ,
\label{kappa}
\end{equation}
where
\begin{equation}
\tilde m_1=\frac{v^2}{M_1} (Y_\nu^\dagger Y_\nu)_{11} \ .
\end{equation}
This expression  is valid for in the region of $M_1\ll 10^{14}$GeV and $\tilde m_1 \geq 10^{-2}$eV \cite{Giudice:2003jh}.
As seen later,  our numerical results are presented under these conditions.
After the reprocessing by sphaleron transitions,
the baryon asymmetry is related to the $B-L$ asymmetry as
\begin{equation}
\eta_B\equiv \frac{n_B}{n_\gamma}=7.04\times \frac{28}{79}Y_{B-L} \ .
\label{etaB}
\end{equation}

In order to estimate the lepton asymmetry $\epsilon_{N_1}$,
 we show the explicit forms given by  the Yukawa matrix elements
  for cases I, II and III as
\begin{align}
{\rm case \ I:}&\quad\frac{{\rm Im}[\{(Y_\nu^\dagger Y_\nu)_{21}\}^2]}{(Y_\nu^\dagger Y_\nu)_{11}}
=\frac{1}{2} f^2(k-1)^2 \sin 2\phi_B \ , \qquad
\tilde m_1 =2 v^2\frac{f^2}{M_0} B^2 \ ,
\label{case1YY} \\
{\rm case\  II:}&\quad\frac{{\rm Im}[\{(Y_\nu^\dagger Y_\nu)_{21}\}^2]}{(Y_\nu^\dagger Y_\nu)_{11}}
= \frac{1}{20} f^2(5k+1)^2 \sin 2\phi_B \ , \ \quad
\tilde m_1=\frac{5}{4}v^2 \frac{f^2}{M_0} \hat B^2 \ ,
\label{case2YY} \\
{\rm case \ III:}&\quad\frac{{\rm Im}[\{(Y_\nu^\dagger Y_\nu)_{21}\}^2]}{(Y_\nu^\dagger Y_\nu)_{11}}
=\frac{1}{20} f^2(k+5)^2 \sin 2\phi_B \ , \qquad
\tilde m_1=\frac{5}{4}v^2 \frac{f^2}{M_0} B^2 \ ,
\label{case3YY}
\end{align}
respectively.
Therefore,  the sign of the  CP asymmetry $\epsilon_{N_1}$ in Eq.~(\ref{asym}) is fixed by the sign of $(-\sin 2\phi_B)$ 
in the I, II, III cases with $M_1\ll M_2$,
while the signs of $Y_{B-L}$ and $\eta_B$ in Eqs.~(\ref{B-L}) and (\ref{etaB}) are given by the sign of  $\sin 2\phi_B$.

It is easily found that $(Y_\nu^\dagger Y_\nu)_{21}$ vanishes for the Yukawa matrices of Eqs.~(\ref{MD23-IH}) and (\ref{MD13}) in subsections 2.2 and 2.3.
Therefore, the  CP asymmetry at the decay of the  right-handed neutrino $N_1$  is not realized for $\rm TM_1$ of IH and  $\rm TM_2$ of both NH and IH.
We present numerical results of the leptogenesis for the three cases of $\rm TM_1$ with NH in the next section.

\section{Correlation between  $\delta_{CP}$	and the lepton asymmetry}

We discuss the correlation between the predicted CP violating phase $\delta_{CP}$
and the observed BAU through the leptogenesis in   cases I,
II and III for $\rm TM_1$ in NH.

In order to determine the free parameter set ($k$, $\phi_B$, $B\sqrt{p}$, $f^2/M_0$) in the neutrino mass matrices in Eqs.~(\ref{case1}), (\ref{case2}) and (\ref{case3}),
we use the result of the global analyses in Refs.~\cite{Esteban:2016qun,deSalas:2017kay} for five data of the mixing angles and neutrino masses~\footnote{
The data of $\sin\theta_{12}$ does not constrain the parameters of the minimal seesaw model for $\rm TM_1$ because the prediction from $\rm TM_1$ is completely consistent with the observed $\sin\theta_{12}$.}.
We can predict the CP violating phase $\delta_{CP}$ by inputting the data within $3\sigma$ in Table 1 \cite{Esteban:2016qun}.
The CP violating measure,  $J_{CP}$ ~\cite{Jarlskog:1985ht} in Eq.~(\ref{J}),  is related to the mixing angles and the CP violating phase as
\begin{equation}
J_{CP}=s_{23}c_{23}s_{12}c_{12}s_{13}c_{13}^2\sin \delta _{CP}~,
\label{Jcp}
\end{equation}
where  $c_{ij}$ ($s_{ij}$) denotes $\cos\theta_{ij}\geq 0 $ ($\sin\theta_{ij} \geq 0$) and $\delta_{CP}$ is the CP violating Dirac phase in the PDG convention \cite{Olive:2016xmw}.
It is noted that the sign of $J_{CP}$ is given by the sign of  $\sin\delta_{CP}$.
\begin{table}[t!]
	\begin{center}
		\begin{tabular}{|c|c|c|}
			\hline
			\  observable \ & $3\sigma$ range  for NH & $3\sigma$ range for IH \\
			\hline
			$|\Delta m_{13}^2|$& \ \   \ \ $(2.407\sim 2.643) \times 10^{-3}{\rm eV}^2$ \ \ \ \
			&\ \ $(2.399\sim 2.635) \times 10^{-3}{\rm eV}^2$ \ \  \\
			\hline
			$\Delta m_{12}^2$&  $(7.03\sim 8.09)  \times 10^{-5}{\rm eV}^2$
			& $(7.03\sim 8.09)  \times 10^{-5}{\rm eV}^2$ \\
			\hline
			$\sin^2\theta_{23}$&  $0.385\sim 0.635$  & $0.393\sim 0.640$ \\
			\hline
			$\sin^2\theta_{12}$& $0.271\sim 0.345$ & $0.271\sim 0.345$ \\
			\hline
			$\sin^2\theta_{13}$&  $0.01934\sim 0.02392$ & $0.01953\sim 0.02408$ \\
			\hline
		\end{tabular}
		\caption{$3\sigma$ range of the global analysis of the neutrino oscillation experimental data
			for NH and IH~\cite{Esteban:2016qun}. }
		\label{tab}
	\end{center}
\end{table}
Although   $J_{CP}$ is defined  in terms of the mixing matrix elements in Eq.~{(\ref{J}),
it is  directly calculated by using the neutrino mass matrices in Eqs.~(\ref{case1}), (\ref{case2}) and (\ref{case3}) for three cases ($\rm TM_1$ in NH).
The CP-odd weak basis invariant ${\cal J}_{CP}$ is given as follows \cite{Branco:2011zb,Castelo-Branco:2014zua}:
\begin{equation}
{ \cal J}_{CP}={\rm Tr}\left [ (M_\nu M_\nu^\dagger)^*, (M_\ell M_\ell^\dagger) \right ]^3
=-6i   \Delta m^6_\ell  \Delta m^6_\nu J_{CP} \ ,
\end{equation}
where $M_\ell$ is the charged lepton mass matrix,  and $\Delta m^6_\ell$
and $\Delta m^6_\nu$ are
\begin{equation}
\Delta m^6_\ell =(m^2_\mu-m^2_e)(m^2_\tau-m^2_\mu)(m^2_\tau-m^2_e) , \quad
\Delta m^6_\nu=(m^2_2-m^2_1)(m^2_3-m^2_2)(m^2_3-m^2_1) ,
\end{equation}
respectively.
By using this formula and the abbreviation $\Delta m_{ij}^2\equiv m_j^2-m_i^2$,
we obtain
\begin{equation}
{\rm case \ I}: \quad
J_{CP}=-\frac{3}{8}\frac{f^{12}}{M_0^6}(B\sqrt{p})^6 (k-1)(k+1)^5 \sin 2\phi_B
\frac{v^{12}}{(\Delta m^2_{13}-\Delta m^2_{12})\Delta m^2_{13}\Delta m^2_{12}}\ ,
\label{JCP-I}
\end{equation}

\begin{equation}
{\rm case \ II}: \quad
J_{CP}=-\frac{3}{32}\frac{f^{12}}{M_0^6}(B\sqrt{p})^6 (5k+1) \sin 2\phi_B
\frac{v^{12}}{(\Delta m^2_{13}-\Delta m^2_{12})\Delta m^2_{13}\Delta m^2_{12}}\ ,
\label{JCP-I}
\end{equation}

\begin{equation}
{\rm case \ III}: \quad
J_{CP}=\frac{3}{32}\frac{f^{12}}{M_0^6}(B\sqrt{p})^6 k^5(k+5) \sin 2\phi_B
\frac{v^{12}}{(\Delta m^2_{13}-\Delta m^2_{12})\Delta m^2_{13}\Delta m^2_{12}}\ .
\label{JCP-I}
\end{equation}
It is remarked that the sign of $J_{CP}$ is classified according to $k$ because of   $\Delta m_{13}^2\gg\Delta m_{12}^2>0$  as follows:
\begin{equation}
 J_{CP}\sim\left\{
 \begin{array}{lclc}
  \sin 2\phi_B \ \ {\rm for } \ -1\leq k\leq 1\ &;&  -\sin 2\phi_B \ \ {\rm for } \ k \leq -1, \  k\geq 1 & \mbox{in case I}\\
 \sin 2\phi_B \ \ {\rm for } \  k\leq -1/5\   \ &;&  -\sin 2\phi_B \ \ {\rm for } \  k \geq -1/5  & \mbox{ in case II }\\
 \sin 2\phi_B \ \ {\rm for } \  k \leq -5, \ k\geq 0  &;& -\sin 2\phi_B \ \ {\rm for } \ -5 \leq k \leq 0  & \mbox{ in case III }
 \end{array}
 \right .  .
 \label{JCP123}
 \end{equation}
It is noted that the sign of $\sin 2\phi_B$ must be positive since the observed $\eta_B$ is positive as discussed below Eq.~(\ref{case3YY}). 
Therefore, we can investigate the sign of $\sin\delta_{CP}$ based on Eqs.~(\ref{Jcp}) and (\ref{JCP123}).
We can predict not only the absolute value of $\delta_{CP}$ but also its sign since the parameter $k$ is constrained by inputting the data of mixing angles and masses in Table 1 \cite{Esteban:2016qun}.

Although the inputting data in Table 1 constrain the parameter set  ($k$, $\phi_B$, $B\sqrt{p}$, $f^2/M_0$),
there are still free parameters, $p=M_2/M_1$ and $M_0=M_2$ to discuss the leptogenesis numerically.
In our calculation, we put $M_0=10^{14}$GeV for convenience.
We will discuss the $M_0$ dependence of our numerical result later.
The parameter $p$ is fixed to reproduce the observed BAU,  $\eta_B=(5.8-6.6)\times 10^{-10}$ $95\%$C.L. \cite{Olive:2016xmw}.
We also take account of that the standard thermal leptogenesis requires $M_1\geq 10^9$GeV\cite{Davidson:2002qv}.
Furthermore, we take $M_1\ll 10^{14}$GeV and $\tilde m_1 \geq 10^{-2}$eV
as we use the approximate form of $\kappa$ in Eq.~(\ref{kappa})~\cite{Giudice:2003jh}.

At first, we discuss the numerical results for case I.
We show the $k$ dependence of the predicted $\delta_{CP}$ by inputting the observed BAU,
$\eta_B=(5.8-6.6)\times 10^{-10}$ $95\%$C.L. \cite{Olive:2016xmw} in Fig.~\ref{fig:case1deltaCP}(a).
As seen in this figure, $\delta_{CP}$ is predicted to be positive  for $-1<k<0$ while it is negative for  $k<-1$ as expected in Eq.~(\ref{JCP123}).
We also show the predicted $\delta_{CP}$ versus $\sin^2 \theta_{23}$  in Fig.~\ref{fig:case1deltaCP}(b).
These predictions are independent of $M_0$ although $M_0=10^{14}$GeV is put in our calculations.
It is concluded that the region of $ k \leq -1$ is favored for the case $M_1 \ll M_2$
if we take account of the recent T2K data \cite{Abe:2017vif,T2K} and the NO$\nu$A data  \cite{Adamson:2017gxd}.
This result is consistent with the Littlest seesaw model with $k=-3$ by King \cite{King:2015dvf}
where the leptogenesis phase yields an observable neutrino oscillation phase with $\delta_{CP}\simeq -\pi/2$.

Let us discuss the case of  $ k \leq -1$ in detail.
The parameters of  $f$ and $p=M_2/M_1$ depend on the magnitude of $M_0$.
As mentioned in Section 2,  $f$ is taken to be real positive.
We show the allowed region on the plane of  $f$ and $p$ for both cases of  $M_0=10^{14}$GeV and  $10^{15}$GeV in Fig. \ref{fig:case1fpm1}(a).
The Yukawa coupling $f$ is predicted in $f<1$, which is preferable in the perturbative calculation of the lepton asymmetry.
For the case of $M_0=10^{15}$GeV, $p$ and $f^2$ are larger
ten times compared to those in $M_0=10^{14}$GeV.
This behavior is easily understood as follows.
The neutrino masses $m_3$, $m_2$ and $\tilde m_1$ are
proportional to $f^2/M_0$ as seen in Eqs.~(\ref{masses}) and (\ref{case1YY})
while the asymmetry parameter $\epsilon_{N_1}$ depends on $f^2/p$
for $p\gg 1$ as seen in  Eqs.~(\ref{approasym}) and (\ref{case1YY}).
Therefore, the numerical result of neutrino masses and the asymmetry  are unchanged if
we enlarge $M_0$, $f^2$ and $p$ ten times simultaneously.

In order to check the availability of the approximate $\kappa$  in Eq.~(\ref{kappa}),
we show  $\tilde m_1$ versus $k$ in Fig.~\ref{fig:case1fpm1}(b).
As seen in this figure, $\tilde m_1$ is $(4.2-7.7)\times 10^{-2}$eV, which satisfies the condition, $\tilde m_1 \geq 10^{-2}$eV in Ref.~\cite{Giudice:2003jh}.
The calculated $\tilde m_1$'s almost overlap for $M_0=10^{14}$GeV and $10^{15}$GeV.

\begin{figure}[t]
	\begin{tabular}{ll}
		\begin{minipage}{0.5\hsize}
			\begin{center}
\includegraphics[{width=\linewidth}]{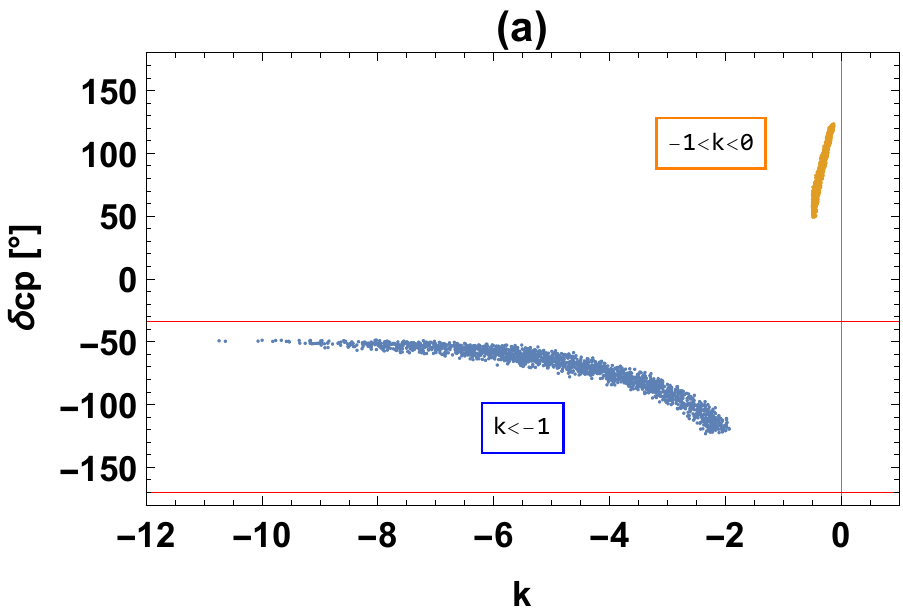}
			\end{center}
		\end{minipage}
		\begin{minipage}{0.5\hsize}
			\begin{center}
\includegraphics[{width=\linewidth}]{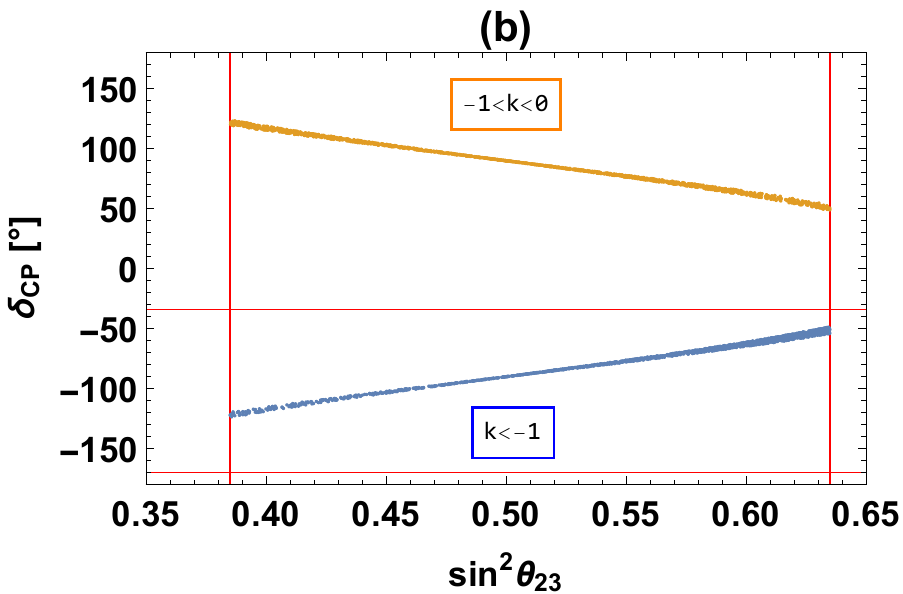}
			\end{center}
		\end{minipage}
	\end{tabular}
	\caption{Predictions in case I.
		The blue  and orange dots denote the region of $k<-1$ and $-1<k<0$, respectively.
		The red lines for $\sin^2\theta_{23}$ and $\delta_{CP}$ denote the experimental bounds of $3\sigma$ (global analyses) and $2\sigma$ (T2K) ranges, respectively:
		  (a)  $\delta_{CP}$ versus $k$,
		(b)   $\delta_{CP}$ versus $\sin^2\theta_{23}$.
	}
	\label{fig:case1deltaCP}
\end{figure}


\begin{figure}[h!]
	\begin{tabular}{ll}
		\begin{minipage}{0.5\hsize}
			\begin{center}
				\includegraphics[{width=\linewidth}]{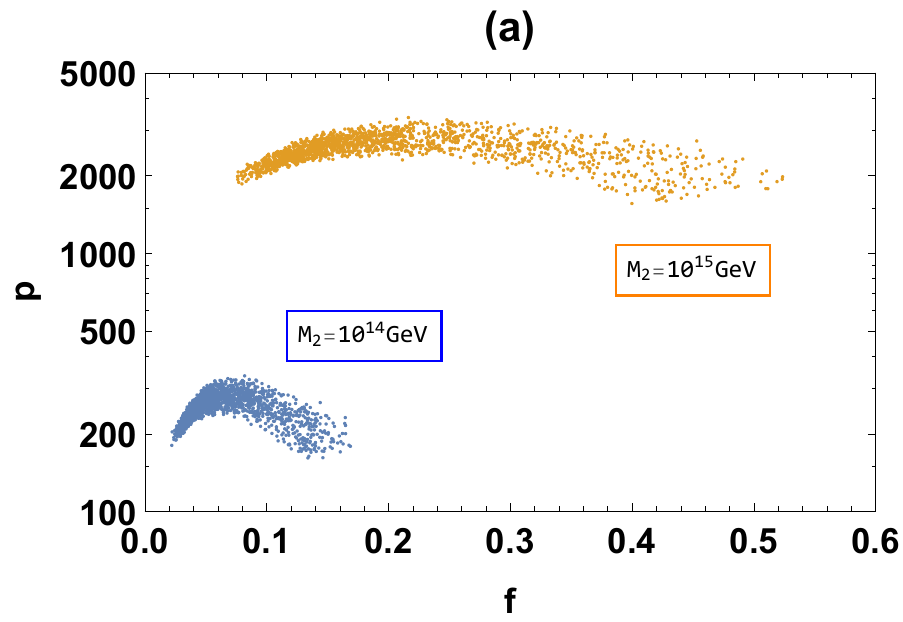}
			\end{center}
		\end{minipage}
		\begin{minipage}{0.5\hsize}
			\begin{center}
				\includegraphics[{,width=\linewidth}]{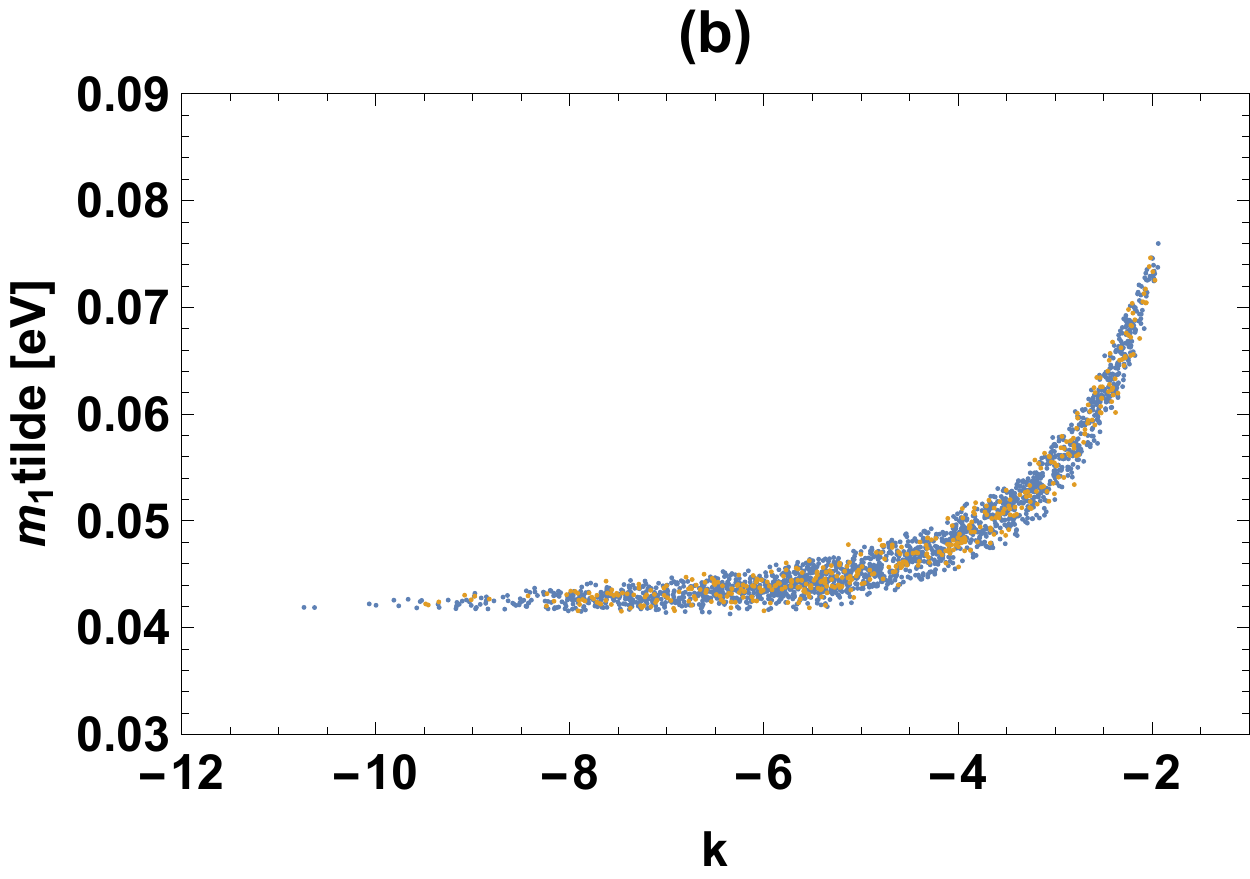}
			\end{center}
		\end{minipage}
	\end{tabular}
	\caption{ Obtained parameters
		in case I for  $M_0=10^{14}$GeV (blue dots) and  $10^{15}$GeV
		(orange dots) with
		$k \leq -1$:  (a)   $p$  versus $f$,
		 (b) $\tilde m_1$ versus $k$.
	}
	\label{fig:case1fpm1}
\end{figure}

\begin{figure}[h!]
	\begin{tabular}{ll}
		\begin{minipage}{0.5\hsize}
			\begin{center}
				\includegraphics[{width=\linewidth}]{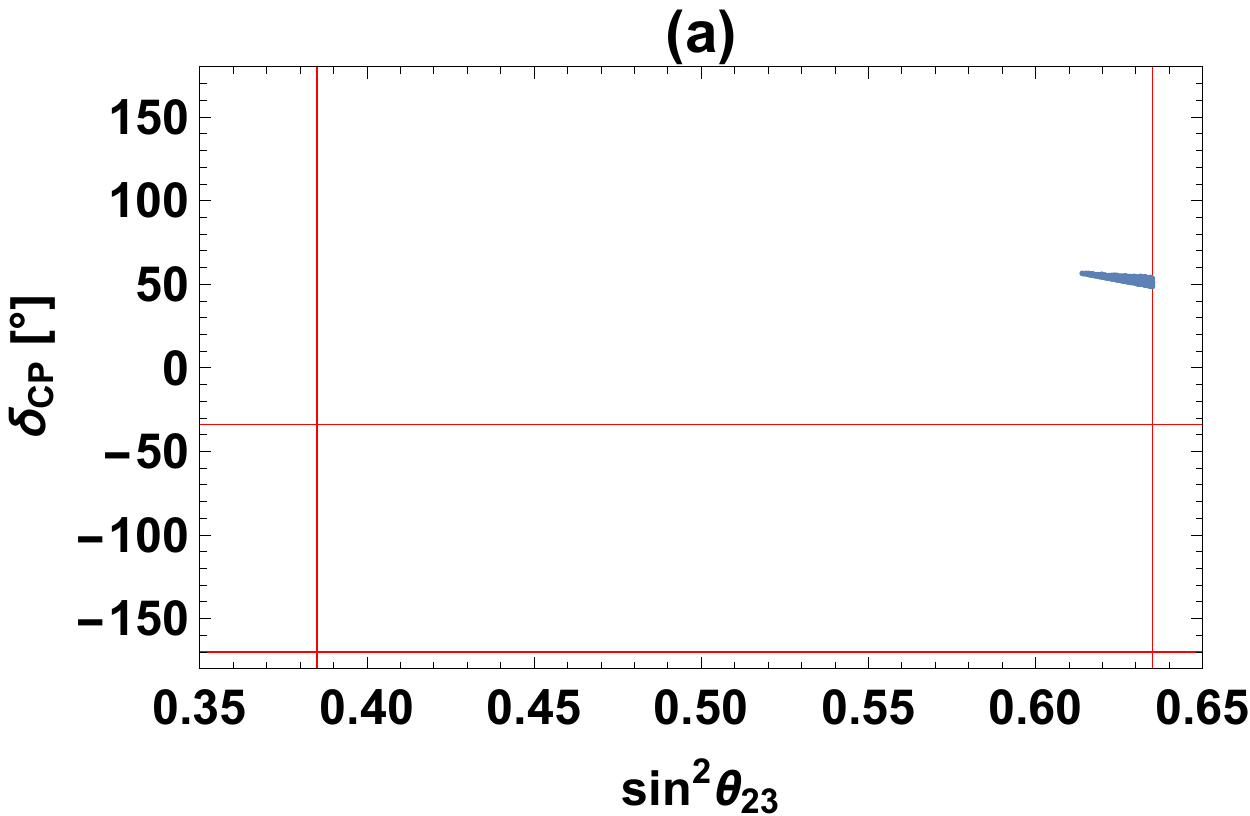}
			\end{center}
		\end{minipage}
		\begin{minipage}{0.5\hsize}
			\begin{center}
				\includegraphics[{width=\linewidth}]{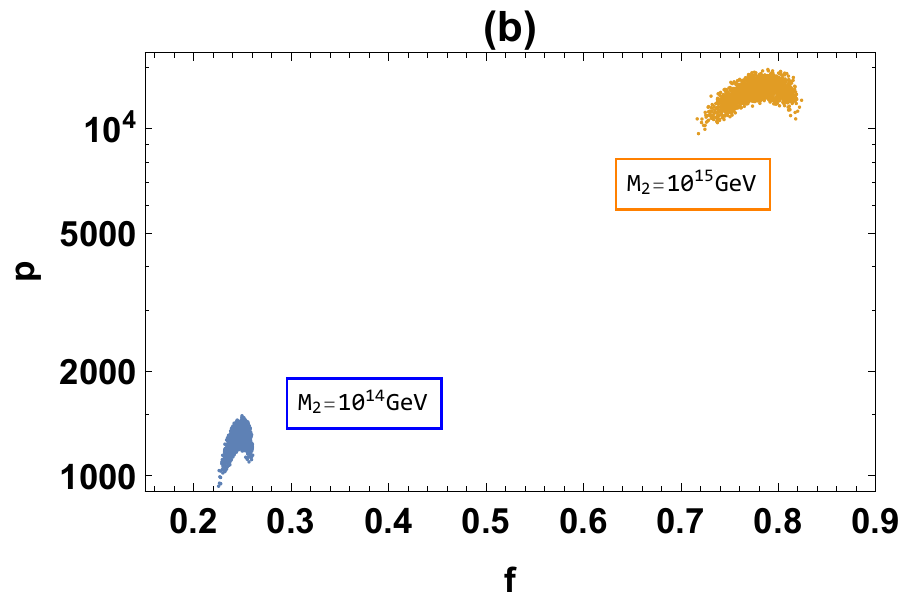}
			\end{center}
		\end{minipage}
	\end{tabular}
	\caption{The predictions in case II.
		(a) The predicted $\delta_{CP}$ versus $\sin^2\theta_{23}$.
		The  red lines for $\sin^2\theta_{23}$ and $\delta_{CP}$ denote the  experimental bounds of $3\sigma$ (global analyses) and $2\sigma$ (T2K) ranges, respectively,
		(b) $p$ versus $f$
		for  $M_0=10^{14}$GeV (blue dots) and  $10^{15}$GeV	(orange dots).
	}
	\label{fig:case2deltaCP}
\end{figure}
Next, we present our result for case II.
In our previous paper~\cite{Shimizu:2017fgu}, it is found
that $k$ is restricted in a narrow range, $k=-1.65\sim -1.10$.
The predicted $\delta_{CP}$ is positive around $+50^\circ$ as seen in Fig. \ref{fig:case2deltaCP}(a) while
the predicted $\sin^2\theta_{23}$ is restricted near the upper bound of the $3\sigma$ range of the experimental data.
The case II is disfavored by the recent T2K and NO$\nu$A data
because the predicted $\delta_{CP}$ to be positive.
We show the allowed region on the plane of  $f$ and $p$ for both cases of  $M_0=10^{14}$GeV and  $10^{15}$GeV in Fig. \ref{fig:case2deltaCP}(b).
The magnitude of $M_0$ is allowed up to $10^{15}$GeV within $f\leq 1$.
We note that  $\tilde m_1$ is   $(1.8-2.4)\times 10^{-2}$eV,
 which  satisfies the condition, $\tilde m_1 \geq10^{-2}$eV in Ref.~\cite{Giudice:2003jh}.

Finally, we discuss the case III.
The allowed range of $k$ is restricted to be very narrow, $k=-0.86\sim -0.71$, as shown in our previous paper \cite{Shimizu:2017fgu}.
We show the predicted $\delta_{CP}$ versus $\sin^2\theta_{23}$ in Fig.~\ref{fig:case3deltaCP}(a), where $\delta _{CP}$ is predicted to be negative around $-125^\circ$.
However, the case III may be soon excluded in the near future as the mixing angle $\sin^2\theta_{23}$ is strongly restricted to the lower bound of the $3\sigma$ range.
In order to compare the allowed parameter regions of $f$ and $p$ in case III with that in cases I and II, we show the plot of  $f$ and $p$ for both cases of  $M_0=10^{14}$GeV and  $10^{15}$GeV in Fig. \ref{fig:case3deltaCP}(b).
The magnitude of  $M_0$ is allowed up to $10^{15}$GeV in the region of 
 $f\leq 1$.
The allowed range of  $\tilde m_1$ is  $(1.9-2.1)\times 10^{-2}$eV,
which satisfies the condition, $\tilde m_1 \geq10^{-2}$eV in Ref.~\cite{Giudice:2003jh}.


\begin{figure}[h!]
	\begin{tabular}{ll}
		\begin{minipage}{0.5\hsize}
			\begin{center}
				\includegraphics[{width=\linewidth}]{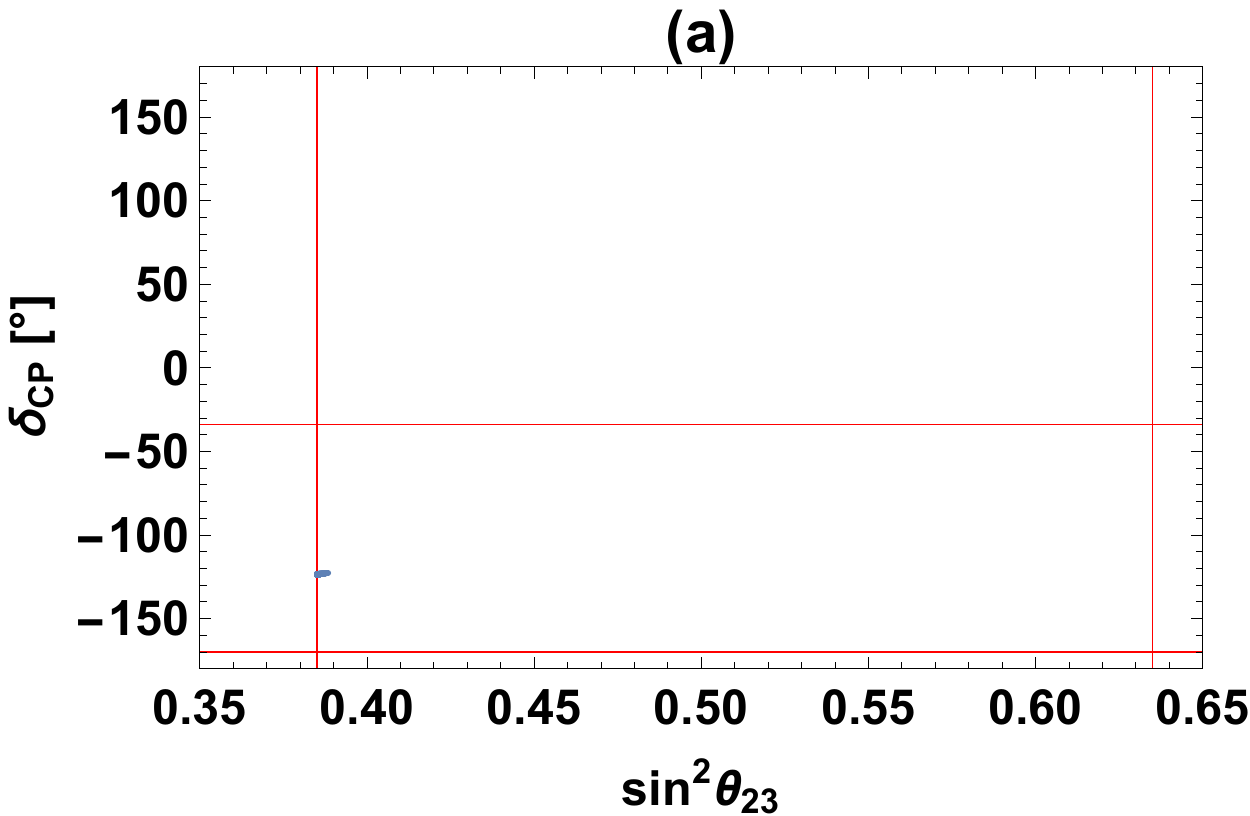}
			\end{center}
		\end{minipage}
		\begin{minipage}{0.5\hsize}
			\begin{center}
				\includegraphics[{width=\linewidth}]{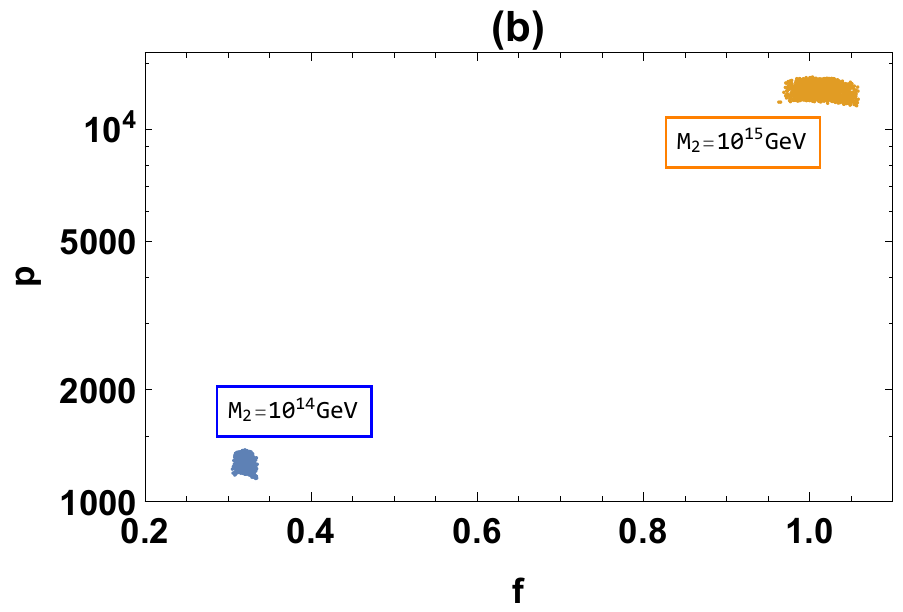}
			\end{center}
		\end{minipage}
	\end{tabular}
	\caption{The prediction in case III.
		(a) The predicted $\delta_{CP}$ versus $\sin^2\theta_{23}$.
		The  red lines for $\sin^2\theta_{23}$ and $\delta_{CP}$ denote the  experimental bounds of $3\sigma$ (global analyses) and $2\sigma$ (T2K) ranges, respectively,
		(b) $p$ versus $f$
		for  $M_0=10^{14}$GeV (blue dots) and  $10^{15}$GeV	(orange dots).
	}
	\label{fig:case3deltaCP}
\end{figure}

\section{Summary and discussions}
We have studied the correlation between the CP violating phase $\delta_{CP}$
and the observed BAU in the minimal seesaw model,
where two right-handed Majorana neutrinos are assumed.
We have also taken the trimaximal mixing pattern for the neutrino flavor ($\rm TM_1$ or $\rm TM_2$) in the diagonal basis of both the charged lepton and right-handed Majorana neutrino mass matrices.
As our model has only one CP violating phase,
we have found the clear correlation between the CP violating phase $\delta_{CP}$ and  BAU for $\rm TM_1$ in NH of neutrino masses.
On the other hand, the lepton asymmetry vanishes for $\rm TM_1$ in IH 
and $\rm TM_2$ in both NH and IH.

We have discussed the three cases of the Dirac neutrino mass matrix for $\rm TM_1$ in NH.
The flavor structure of the Dirac neutrino mass matrix is fixed for case I.
The parameter $k$ should be smaller than $-1$ in order to predict a negative  $\delta_{CP}$,
which is indicated by the recent T2K and NO$\nu$A data.
Our result is completely consistent with the Littlest seesaw model \cite{King:2015dvf} which is the specific model of case I with $k=-3$.
It is emphasized that our Dirac neutrino mass matrix predicts the negative sign of $\delta_{CP}$ and the observed value of BAU as far as we take  $k<-1$ under the condition, $M_1\ll M_2$.

In our calculation, we have assumed $M_1\ll M_2$ for the right-handed Majorana neutrino masses.
For the case of  $M_1\gg M_2$,
one can obtain the numerical result by switching  $M_1$ and $M_2$ with 
the labels $1$ and $2$ in Eqs.~(\ref{case1YY}), (\ref{case2YY}) and (\ref{case3YY}).
Note that the left-handed neutrino mass matrix in Eq.~(\ref{TM1NH}) is unchanged for $M_0=M_1$ and $p=M_1/M_2$.
Therefore, the relevant  lepton asymmetry $\epsilon_{N_2}$ is given by 
$ {\rm Im}[\{(Y_\nu^\dagger Y_\nu)_{12}\}^2]$
where the Yukawa matrix $Y_\nu$ is expressed as Eqs.~(\ref{MD1}),  (\ref{MD2}) or (\ref{MD3}).
Since $(Y_\nu^\dagger Y_\nu)_{12}$ is the complex conjugate of $(Y_\nu^\dagger Y_\nu)_{21}$,
we get the opposite sign of the CP violating phase $\delta_{CP}$
 compared with the case of $M_1\ll M_2$.
Namely, the case with $-1<k<0$ is preferred in contrast to our result of $k<-1$ in case I.
It is also noticed that the case II is favored while the case III is disfavored for  $M_1\gg M_2$.
The mass hierarchy between  $M_1$ and $M_2$ is an important key
parameter  to obtain a robust prediction of the correlation between  $\delta_{CP}$ and BAU.
It may be interesting that the   Froggatt-Nielsen mechanism  \cite{Froggatt:1978nt} 
 is added to our minimal seesaw model to fix the $M_2/M_1$ ratio.
 
Finally, we add a comment.
Since our Dirac neutrino mass matrices are given at the high energy scale
 to predict the BAU,
we should examine the renormalization group correction for the neutrino mixing matrix. However, it is very small since the lightest neutrino mass $m_1$ vanishes in our model  as seen in Ref.~\cite{Gehrlein:2016fms}.

\vspace{1 cm}
\noindent
{\bf Acknowledgment}

This work is supported by JSPS Grants-in-Aid for Scientific Research
 16J05332 (YS) and 15K05045, 16H00862 (MT).

\appendix
\section*{Appendix}
\section{Minimal seesaw mass matrix}


We study the minimal seesaw model in the basis of the diagonal $2\times 2$ right-handed Majorana neutrino mass matrix $M_R$ given in general as:
\begin{equation}
M_R=\begin{pmatrix}
M_1 & 0 \\
0 & M_2
\end{pmatrix}
=M_0
\begin{pmatrix}
p^{-1} & 0 \\
0 & 1
\end{pmatrix},
\end{equation}
where $v=174.1$GeV, $M_0$ is the mass scale of the right-handed Majorana neutrino and $p$ is the ratio  $M_2/M_1$.
The relevant Dirac neutrino mass matrix $M_D$ is defined as
\begin{equation}
M_D=v Y_\nu=v
\begin{pmatrix}
a & d \\
b & e \\
c & f
\end{pmatrix},
\end{equation}
where $a\sim f$ are complex parameters.
By using the type I seesaw mechanism, the left-handed Majorana neutrino mass matrix $M_{\nu }$ is given by
\begin{equation}
M_{\nu }= -M_DM_R^{-1}M_D^T= - \frac{ v^2}{M_0}
\begin{pmatrix}
a^2p+d^2 & abp+de & acp+df \\
abp+de & b^2p+e^2 & bcp+ef \\
acp+df & bcp+ef & c^2p+f^2
\end{pmatrix}.
\label{Aleft-handed-Majorana-1}
\end{equation}
By turning the neutrino mass matrix $M_\nu$ to the TBM mixing basis, $M_{\nu }$ is given as:
\begin{align}
\hat M_\nu\equiv  V_{\text{TBM}}^T M_\nu V_{\text{TBM}}= - \frac{v^2}{M_0}
\begin{pmatrix}
\frac{A_\nu ^2p+D_\nu^2}{6} & \frac{A_\nu B_\nu p+D_\nu  E_\nu}{3\sqrt{2}} &
\frac{A_\nu C_\nu p+D_\nu F_\nu}{2\sqrt{3}} \\
\frac{A_\nu B_\nu p+D_\nu  E_\nu}{3\sqrt{2}} & \frac{B_\nu^2p+E_\nu^2}{3} &
\frac{B_\nu C_\nu p+E_\nu F_\nu}{\sqrt{6}} \\
\frac{A_\nu C_\nu p+D_\nu F_\nu}{2\sqrt{3}} &\frac{B_\nu C_\nu p+E_\nu F_\nu}{\sqrt{6}} & \frac{C_\nu^2p+F_\nu^2}{2}
\end{pmatrix},
\label{Aleft-handed-Majorana-TBM}
\end{align}
where
\begin{align}
&& A_\nu\equiv 2a-b-c, \qquad B_\nu\equiv a+b+c, \qquad C_\nu\equiv c-b, \nonumber\\
&& D_\nu\equiv 2d-e-f, \qquad E_\nu\equiv d+e+f, \qquad F_\nu\equiv f-e.
\end{align}
Based on these formulae,
we discuss the additional $2-3$ family rotation ($\rm TM_1$) and $1-3$ family rotation ($\rm TM_2$)
to the TBM mixing for both NH and IH in the following subsections.

\subsection{$\rm TM_1$: $2-3$ family  rotation in NH}
We consider the case of NH in $\rm TM_1$.
Since $(1,1)$, $(1,2)$, $(2,1)$, $(1,3)$, and $(3,1)$ entries of the matrix must be zero,
conditions for the additional $2-3$ rotation to the TBM mixing are
\begin{equation}
A_\nu=2a-b-c=0 \ ,\qquad D_\nu=2d-e-f=0 \ ,
\end{equation}
where $(a,b,c)$ are supposed to be independent of $(d,e,f)$.
Under these conditions, the mass matrix Eq.~(\ref{Aleft-handed-Majorana-TBM}) is rewritten as
\begin{equation}
\hat M_\nu = -\frac{v^2}{M_0}
\begin{pmatrix}
0 & 0 & 0 \\
0 & \frac{3}{4}\left ((b+c)^2p+(e+f)^2\right ) &
\frac{1}{2}\sqrt{\frac{3}{2}}\left ((c^2-b^2)p-e^2+f^2\right ) \\
0 & \frac{1}{2}\sqrt{\frac{3}{2}}\left ((c^2-b^2)p-e^2+f^2\right ) &
\frac{1}{2}\left ((b-c)^2p+(e-f)^2\right )
\end{pmatrix},
\label{Aleft-handed-Majorana-general}
\end{equation}
where the lightest neutrino mass $m_1$ is zero.
The  Dirac neutrino mass matrix is  given  as:
\begin{equation}
M_D=v Y_\nu=v
\begin{pmatrix}
\frac{b+c}{2} & \frac{e+f}{2} \\
b & e \\
c & f
\end{pmatrix}.
\label{Ageneral-texture}
\end{equation}
We discuss the specific three cases from this texture.
Putting one zero in the texture Eq.~(\ref{Ageneral-texture}),
we have three possible patterns as follows:
\begin{equation}
\mbox{(I) $b+c=0$,\qquad (I\hspace{-.1em}I) $c=0$, \qquad
	(I\hspace{-.1em}I\hspace{-.1em}I) $b=0$.}
\end{equation}
Corresponding Dirac neutrino mass matrices are
\begin{equation}
M_D=\left\{
\begin{array}{cl}
v\begin{pmatrix}
0 & \frac{e+f}{2} \\
b & e \\
-b & f
\end{pmatrix} & \mbox{ for (I) $b+c=0$}\\
v\begin{pmatrix}
\frac{b}{2} & \frac{e+f}{2} \\
b & e \\
0 & f
\end{pmatrix} & \mbox{ for (I\hspace{-.1em}I) $c=0$}\\
v\begin{pmatrix}
\frac{c}{2} & \frac{e+f}{2} \\
0 & e \\
c & f
\end{pmatrix} & \mbox{ for (I\hspace{-.1em}I\hspace{-.1em}I) $b=0$}
\end{array}
\right . .\label{AMD1}
\end{equation}
We get another set by switching  the first and second columns of the Dirac neutrino mass matrix in Eq.~(\ref{AMD1}).
However, the neutrino mass matrix $\hat M_\nu$ is unchanged 
under the rescale of parameters.
Therefore, we show only three cases in  Eq.~(\ref{AMD1}).
It is noted that the switching  the first and second columns of the Dirac neutrino mass matrix leads to the change of the sign for the leptogenesis.

We show the neutrino mass matrix $\hat M_\nu$ for three cases:
\begin{equation}
{\rm Case\ I}\ : \
\hat M_{\nu }= - \frac{v^2}{M_0}
\begin{pmatrix}
0 & 0 & 0 \\
0 & \frac{3}{4}(e+f)^2 & -\frac{1}{2}\sqrt{\frac{3}{2}}(e-f)(e+f) \\
0 & -\frac{1}{2}\sqrt{\frac{3}{2}}(e-f)(e+f) & 2b^2 p+\frac{1}{2}(e-f)^2
\end{pmatrix}
,
\end{equation}

\begin{equation}
{\rm Case\ II}\ : \
\hat M_{\nu }= - \frac{v^2}{M_0}
\begin{pmatrix}
0 & 0 & 0 \\
0 & \frac{3}{4}[b^2 p+(e+f)^2] & -\frac{1}{2}\sqrt{\frac{3}{2}}[b^2 p+(e-f)(e+f)] \\
0 &  -\frac{1}{2}\sqrt{\frac{3}{2}}[b^2 p +(e-f)(e+f)] & \frac{1}{2}[b^2 p +(e-f)^2]
\end{pmatrix}
,
\end{equation}

\begin{equation}
{\rm Case\ III}\ : \
\hat M_{\nu }= - \frac{v^2}{M_0}
\begin{pmatrix}
0 & 0 & 0 \\
0 & \frac{3}{4}[c^2 p+(e+f)^2] & -\frac{1}{2}\sqrt{\frac{3}{2}}[-c^2 p +(e-f)(e+f)] \\
0 &  -\frac{1}{2}\sqrt{\frac{3}{2}}[-c^2 p+(e-f)(e+f)] & \frac{1}{2}[c^2 p+(e-f)^2]
\end{pmatrix}
.
\end{equation}

\subsection{$\rm TM_1$:  $2-3$ rotation in IH}
We discuss the case of IH in $\rm TM_1$.
In order to give the additional $2-3$ family rotation to the TBM mixing,
the $(1,2)$, $(1,3)$, $(2,1)$, and $(3,1)$ elements in Eq.~(\ref{Aleft-handed-Majorana-TBM}) should vanish.
These conditions lead to
\begin{equation}
A_\nu=a+b+c=0,\qquad C_\nu=c-b=0,\qquad D_\nu=2d-e-f=0.
\end{equation}
The Dirac neutrino mass matrix is
\begin{equation}
 M_D=v Y_\nu=v
\begin{pmatrix}
-2b & \frac{e+f}{2} \\
b & e \\
b& f
\end{pmatrix}.
\label{Ageneral-texture-inverted}
\end{equation}
The neutrino mass matrix $\hat M_\nu$ is given as:
\begin{equation}
\hat M_{\nu }= - \frac{v^2}{M_0}
\begin{pmatrix}
6b^2 p & 0 & 0 \\
0 & \frac{3}{4}(e+f)^2 & -\frac{1}{2}\sqrt{\frac{3}{2}}(e-f)(e+f) \\
0 & -\frac{1}{2}\sqrt{\frac{3}{2}}(e-f)(e+f) & \frac{1}{2}(e-f)^2
\end{pmatrix}
,
\end{equation}
where the neutrino mass $m_3$ vanishes.
\subsection{$\rm TM_2$:  $1-3$ rotation in NH or IH}

We discuss the case of the additional $1-3$ family rotation to the TBM mixing.
This case is called as $\rm TM_2$.
Then, $(1,2)$, $(2,3)$, $(2,1)$, and $(3,2)$ elements in Eq.~(\ref{Aleft-handed-Majorana-TBM}) should vanish.
These conditions lead to
\begin{equation}
A_\nu=2a-b-c=0, \qquad C_\nu=c-b=0, \qquad E_\nu=d+e+f=0\ .
\end{equation}
The  Dirac neutrino mass matrices is
\begin{equation}
M_D=v Y_\nu=v
\begin{pmatrix}
b & -e-f \\
b & e \\
b& f
\end{pmatrix},
\label{Ageneral-texture-inverted}
\end{equation}
and the neutrino mass matrix $\hat M_\nu$ is given  as
\begin{equation}
\hat M_\nu = - \frac{v^2}{M_0}
\begin{pmatrix}
\frac{3}{2}(e+f)^2 & 0 &\frac{\sqrt{3}}{2}(e^2-f^2)\\
0 & 3b^2 p & 0 \\
\frac{\sqrt{3}}{2}(e^2-f^2) &0 & \frac{1}{2}(e-f)^2
\end{pmatrix} .
\label{Aleft-handed-Majorana-general-inverted}
\end{equation}
The mass eigenvalue $m_1$ or $m_3$ vanishes for NH or IH, respectively.

\section{ $2-3$ flavor mixing ${\cal V}$ }

The explicit forms of the $2-3$ flavor mixing ${\cal V}$
in Eq.~(\ref{rotaion23})  are given as follows.

\noindent
For case I:
\begin{equation}
{\cal V}=-\frac{f^2 v^4}{M_0^2}
\frac{\sqrt{6}(k^2-1)[ (5k^2+2k+5)+8B^2p\ e^{2i\phi_B}]}
{16 m_3^2+3\frac{f^4v^4}{M_0^2}(k+1)^2(5k^2+2k+5)} \ .
\end{equation}
For case II:
\begin{equation}
{\cal V}=
-\frac{f^2 v^4}{M_0^2}\frac
{\sqrt{6}\left[(k^2-1)(5k^2+2k+5)+5B^4p^2+2B^2p(5k+1)(k\cos{2\phi_B}+i\sin{2\phi_B})\right]}
{16m_3^2-3\frac{f^4v^4}{M_0^2}\left[(k+1)^2(5k^2+2k+5)+5B^4p^2+2B^2p(k+1)(5k+1)\cos{2\phi_B}\right]}.
\end{equation}
For case III:
\begin{equation}
{\cal V}=
-\frac{f^2 v^4}{M_0^2}\frac
{\sqrt{6}\left[(k^2-1)(5k^2+2k+5)-5B^4p^2-2B^2p(k+5)(\cos{2\phi_B}+ik\sin{2\phi_B})\right]}
{16m_3^2-3\frac{f^4v^4}{M_0^2}\left[(k+1)^2(5k^2+2k+5)+5B^4p^2+2B^2p(k+1)(k+5)\cos{2\phi_B}\right]}.
\end{equation}


\end{document}